\documentclass[usenatbib]{mn2e}
\usepackage{ulem}
\usepackage{amsmath}
\usepackage{graphicx}
\usepackage{subfigure}
\usepackage{multirow}
\usepackage{color}
\usepackage{enumitem}
\begin{document}
\title[Equipartition in CMZs]{Equipartition and Cosmic Ray Energy Densities in Central Molecular Zones of Starbursts}

\author[Yoast-Hull, Gallagher, \& Zweibel]{Tova M. Yoast-Hull$^{1,2}$\thanks{E-mail: yoasthull@wisc.edu}, John S. Gallagher III$^3$, and Ellen G. Zweibel$^{1,2,3}$\\
$^1$Department of Physics, University of Wisconsin-Madison, WI, USA, 53706\\
$^2$Center for Magnetic Self-Organization in Laboratory and Astrophysical Plasmas, University of Wisconsin-Madison, WI, USA, 53706\\
$^3$Department of Astronomy, University of Wisconsin-Madison, WI, USA, 53706}

\maketitle


\begin{abstract}
The energy densities in magnetic fields and cosmic rays (CRs) in galaxies are often assumed to be in equipartition, allowing for an indirect estimate of the magnetic field strength from the observed radio synchrotron spectrum.  However, both primary and secondary CRs contribute to the synchrotron spectrum, and the CR electrons also loose energy via bremsstrahlung and inverse Compton.  While classical equipartition formulae avoid these intricacies, there have been recent revisions that account for the extreme conditions in starbursts.  Yet, the application of the equipartition formula to starburst environments also presupposes that timescales are long enough to reach equilibrium.  Here, we test equipartition in the central molecular zones (CMZs) of nearby starburst galaxies by modeling the observed $\gamma$-ray spectra, which provide a direct measure of the CR energy density, and the radio spectra, which provide a probe of the magnetic field strength. We find that in starbursts, the magnetic field energy density is significantly larger than the CR energy density, demonstrating that the equipartition argument is frequently invalid for CMZs.
\end{abstract}


\section{Introduction}

This paper presents the results from an empirical study of energy densities in the interstellar media of galaxies covering a range of star formation rate (SFR) intensities. In the solar neighborhood, the energy densities in cosmic rays ($U_{CR}$), the radiation field ($U_{\text{rad}}$), thermal gas ($U_{g}$), and the magnetic field ($U_{B}$) are approximately equal at the level of $\sim 1$~eV~cm$^{-3}$.  This situation of near energy equipartition is likely to be the result of the effects of feedback which sets both the SFR and defines the structure of the interstellar medium (ISM).  As such, the assumption of equipartition is often used in systems similar to the Milky Way to derive galactic energy densities from the observed radio continuum.

In addition to the equipartition assumption, a related approach is to derive $U_{B}$ from the minimum energy assumption for radio synchrotron emission. The ``classical'' formula adopted for minimum energy relates the energy densities such that $U_{B} \approx 0.75 ~ U_{CR}$, when the synchrotron spectral index is $\sim 0.8 - 1.0$ \citep{Beck05}.  Thus, the underlying assumptions of both methods imply that $U_{B} \sim U_{CR}$ and that leptonic ($e^{\pm}$) cosmic rays (CRs) mainly lose energy via synchrotron radiation \citep[e.g.,][]{Schober15}.  Recent revisions to the original formula take into account additional energy losses from inverse Compton cooling and bremsstrahlung and the production of secondary $e^{\pm}$ CRs, both of which are significant in dense environments with intense radiation fields \citep[e.g.,][]{Torres04,Beck05,delPozo09,Lacki13a,Persic14}.  These additional losses, in combination with galactic winds, reduce the energy density in leptonic CRs such that an increase in $U_{B}$ is needed to produce sufficient synchrotron radiation. The observed correlation between far infrared and radio luminosities thus requires that $U_{B}$ scales with $U_{\text{rad}}$ so as to preserve the synchrotron luminosity in systems with intense radiation fields \citep{Volk89,Murphy09,Lacki10}.

%
\begin{table*}
\begin{minipage}{\textwidth}
\centering
\scriptsize
\caption{Galaxy Properties}
\begin{tabular}{lcccccc}
\hline
 & Supernova & (CMZ) & Molecular & Acceleration & Spectral & \\
 & Power & Volume & Gas Mass & Efficiency & Index & Ref.\\
 & (erg~yr$^{-1}$) & ($10^{6}$ pc$^{3}$) & ($10^{8} M_{\odot}$) & ($\eta$) & ($\Gamma$) & \\
\hline
Milky Way & $2 \times 10^{48}$ & $1.2 \times 10^{5}$ & 20 & 0.1 & 2.7 & \\
M31 & $\sim 2 \times 10^{48}$ & $2.8 \times 10^{4}$ & 4 & 0.1 & 2.56 & 1,2\\
LMC & $3.5 \times 10^{47}$ & 2800 & 0.4--0.7 & 0.1 & 2.26 & 3,4\\
\hline
M82 & $7 \times 10^{48}$ & 25 & 4 & 0.2 & 2.2 -- 2.3 & 5\\
NGC 253 & $10^{49}$ & 7.1 & 3 & 0.05 & 2.3 & 6\\
\hline
Arp 220 East & $7 \times 10^{49}$ & 1.2 & 6 & 0.05 -- 0.2 & 2.1 -- 2.3 & 7\\
Arp 220 ST & $7 \times 10^{49}$ & 1.8 & 4 & 0.05 -- 0.1 & 2.2 -- 2.3 & 7\\
Arp 220 CND & $1.3 \times 10^{50}$ & 0.23 & 6 & 0.05 -- 0.1 & 2.2 -- 2.3 & 7\\
\hline
\multicolumn{7}{l}{References -- (1) \citet{Pavlidou01}; (2) \citet{Nieten06}; (3) \citet{Maoz10};}\\
\multicolumn{7}{l}{(4) \citet{Fukui99}; see (5) YEGZ; (6) \citet{YoastHull14}; (7) \citet{YoastHull15}.}\\
%
%
\end{tabular}
\end{minipage}
\end{table*}

In this paper, we consider the impact of departures from the equilibrium condition on interstellar energy densities in the central molecular zones (CMZs) of starburst galaxies. We utilize the YEGZ models \citep{YoastHull13,YoastHull14,YoastHull15} which assume steady state conditions and are based on observed SFRs and supernova rates to derive $U_{CR}$ and $U_{B}$ in starbursts. We demonstrate the importance of $\gamma$-ray measurements in narrowing the range of possible models by showing how a wide variety of parameters can fit radio synchrotron observations alone when the equipartition assumption is relaxed. Our work complements the parametric investigation of energy density scalings by \citet{Lacki10, Schleicher13}. In the former studies, the SFR is assumed to follow from the gas surface density $\Sigma_{g}$ through the Schmidt-Kennicutt relationship. This correlation between the SFR and $\Sigma_{g}$, however, breaks down under extreme conditions, such as those found in the Galactic Center \citep[e.g.,][]{Longmore13, Xu15}.  We argue that these are the types of environments where equipartition and equilibrium assumptions also are likely to fail.

\section{Model and Results}

The relationship between CRs and the observed non-thermal emission from galaxies is complex.  CRs both hadronic and leptonic, primary and secondary, interact with molecular gas, magnetic fields, and radiation fields to produce non-thermal synchrotron emission, high-energy $\gamma$-rays, and neutrinos.  Thus, to accurately constrain unknown quantities such as $U_{\text{CR}}$ and $U_{B}$ and to assess the equipartition relationship in galaxies, it is critical to model all of the physical processes involved and to have enough different types of observations to derive the unknown quantities independently.  Here, we compare the predicted radio and $\gamma$-ray spectra with observations to constrain $U_{B}$ and $U_{\text{CR}}$, respectively, using our semi-analytic, single-zone YEGZ model for CR interactions.

\subsection{YEGZ Models}

The YEGZ model is specifically designed for CMZs in starburst systems, regions of $\sim 200 - 500$ pc, in which interactions with the starburst environment and advective losses via a galactic wind are strong enough that energy-dependent diffusion can be considered negligible.  The spectra and lifetimes for primary and secondary CRs are computed based on a variety of observational data including (see Table 1): the supernova rate ($\nu_{\text{SN}}$), which provides a measure of the total energy input into CRs; the molecular gas mass ($M_{\text{mol}}$), which is used to compute the average gas density ($n_{\text{ISM}}$) sampled by the CRs; the far-infrared luminosity ($L_{\text{IR}}$) and dust temperature ($T_{\text{dust}}$), which are used to model the radiation field; the galaxy distance and inclination; and the starburst region volume.  An important feature of all models for CR interactions is the dependence on volume rather than surface densities as these determine the relevant interaction rates \citep[e.g.,][]{Niklas97, Schleicher13}.

\begin{figure}
\centering
 \subfigure[CR Electron Energy Loss Rate]{
  \includegraphics[width=0.65\linewidth]{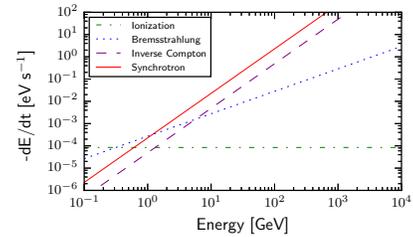}}
 \subfigure[CR Lifetimes]{
  \includegraphics[width=0.65\linewidth]{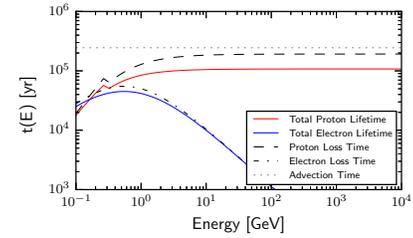}}
\caption{Best-fitting YEGZ models for M82, with $\eta \sim 0.2$, $p \sim 2.3$, $B \sim 300 ~ \mu$G, $v_{\text{adv}} \sim 400$ km~s$^{-1}$, $U_{\text{rad}} \sim 480$ eV~cm$^{-3}$.}
\end{figure}
\begin{figure*}
 \subfigure[$U_{B}$ vs $U_{CR}$]{
  \includegraphics[width=0.25\linewidth]{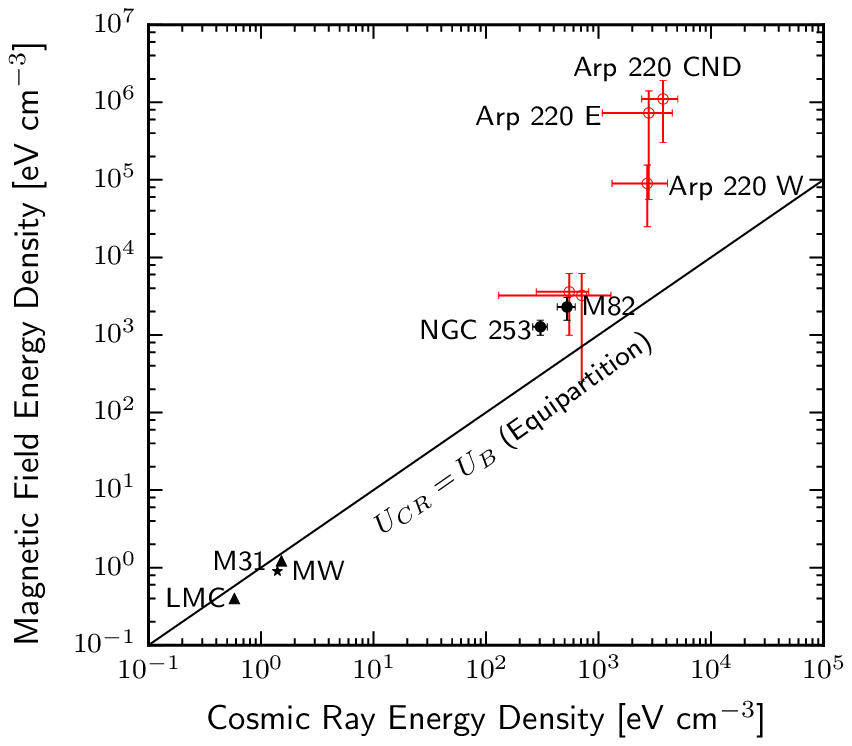}}
 \subfigure[$U_{B}$ vs $U_{rad}$]{
  \includegraphics[width=0.25\linewidth]{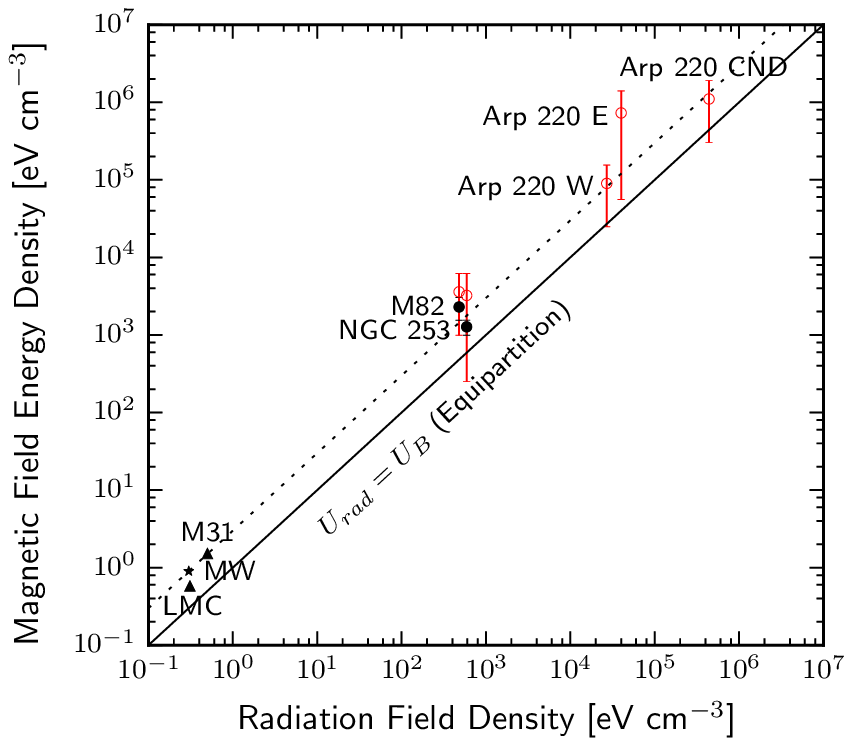}}
 \subfigure[$U_{B} / U_{CR}$ vs $n_{\text{ISM}}$]{
  \includegraphics[width=0.25\linewidth]{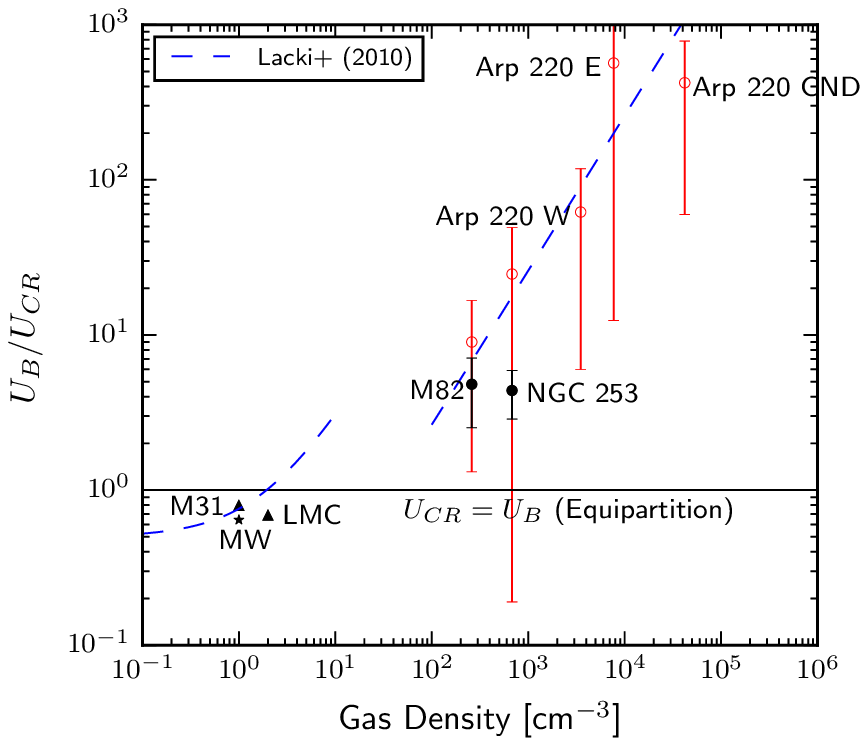}}
\caption{$U_{B}$, $U_{\text{CR}}$, $U_{\text{rad}}$ are shown for all star-forming galaxies.  In the center panel, the dotted line denotes $U_{B} = 3 U_{\text{rad}}$, consistent with \citet{Murphy09}.  The dashed blue line shown on the right is from \citet{Lacki10}, see Figure 15, and is of the form $U_{B} / U_{CR} \propto \Sigma_{g}$.  The discontinuity in the line is due to the use of $\Sigma_{g}$ instead of $n_{\text{ISM}}$ by \citet{Lacki10}.  For each panel, galaxies for which both radio and $\gamma$-ray data was analyzed are shown with black, filled symbols and galaxies with only radio data analysis are denoted with red, open symbols.  Starburst galaxies are denoted with circles, M31 \& the LMC with triangles, and the Milky Way is denoted with a star.}
\end{figure*}

Our YEGZ models assume a power-law source function normalized to the energy input into CRs per volume.  The total energy input into CRs is set as the product of the supernova rate ($\nu_{\text{SN}}$), the typical supernova explosion energy ($10^{51}$ erg), and the acceleration efficiency ($\eta \sim 0.1$, the fraction of the supernova energy which goes into CRs).  We assume a primary proton to electron ratio of 50, and inclusion of secondaries puts the total ratio closer to 25, though this varies with energy loss rate and advection timescale.  The energy loss timescale includes losses from ionization, pion production, bremsstrahlung, inverse Compton, and synchrotron (see Figure 1).  For advection, we leave the wind speed as a free parameter, and while this significantly impacts the CR protons, it has little impact on the CR electrons.

From the total CR populations, we calculate the radio and $\gamma$-ray spectra.  Using $\chi^{2}$ tests to compare the models with the observed spectra, we can constrain the spectral index of the source function ($p$), wind speed ($v_{\text{adv}}$), magnetic field strength ($B$), ionized gas density ($n_{\text{ion}}$), and absorption fraction ($f_{\text{abs}}$).  Wind speed, which sets the total CR proton lifetime, and spectral index are best constrained by the $\gamma$-ray spectrum, and magnetic field strength and the thermal radio contribution, set by the ionized gas density and absorption fraction (fraction of synchrotron emission absorbed in ionized gas, see \cite{YoastHull14} for details), are best measured with the radio spectrum.  Thus, the $\gamma$-ray and radio spectra act as separate probes of the CR proton and CR electron populations and thus $U_{\text{CR}}$ and $U_{B}$.


\subsection{Particle and Field Energy Densities}

For each galaxy, we calculated the reduced $\chi^{2}$ value for a given set of parameters, with separate $\overline{\chi^{2}}$ values for the radio ($\overline{\chi^{2}_{R}}$) and $\gamma$-ray ($\overline{\chi^{2}_{\gamma}}$) spectra.  To determine the best-fitting joint solution, we combined the reduced $\chi^{2}$ values, equally weighting the individual spectra such that $\overline{\chi^{2}}_{tot} = (\overline{\chi^{2}_{R}} + \overline{\chi^{2}_{\gamma}}) / 2$.  Models within $3\sigma$ of the best-fitting model were considered acceptable models.  We then calculated $U_{\text{CR}}$ and $U_{B}$ for each acceptable model (see Figure 2).

\subsubsection{Starburst Galaxies: M82, NGC 253, Arp 220}

Using the updated version of the YEGZ models, we have re-worked our analysis of the radio and $\gamma$-ray spectra for M82 and NGC~253 and have also applied the models to the radio spectrum for the nuclei of Arp 220.  Our previous results for M82 and NGC~253 are consistent with the newer models, where the best-fitting magnetic field strengths range from 200--250  $\mu$G (NGC~253) and 250--350 $\mu$G (M82) \citep{YoastHull13,YoastHull14}.  The starburst nuclei of Arp 220 present a more extreme case where milligauss strength magnetic fields (4.0--7.5 mG in the eastern nucleus and 1.0--2.5 mG in the western nucleus) are necessary to overcome both intense radiation fields and large amounts of dense molecular gas \citep{YoastHull15} and are consistent with direct measurements of the magnetic field on small spatial scales \citep{Barcos15,McBride15}.

Though the observed $\gamma$-ray spectra for M82 and NGC~253 are quite similar, we find that a slightly larger acceleration efficiency ($\eta = 0.2$; or equivalently a somewhat higher supernova rate) is required for M82 and a slightly smaller acceleration efficiency ($\eta = 0.05$) is necessary to fit NGC~253.  Thus, for M82 $U_{\text{CR}}$ ranges from 420 to 620 eV~cm$^{-3}$ and from 260 to 350 eV~cm$^{-3}$ for NGC~253 (see Table 2, Figure 2).  In Arp 220, acceleration efficiencies from $\eta = 0.05-0.2$ produce acceptable fits to the radio spectra, resulting in a larger span in the values of $U_{\text{CR}}$ from 1080 to 4520 eV~cm$^{-3}$ in the eastern nucleus, 1320 to 4110 eV~cm$^{-3}$ in the surrounding torus of the western nucleus, and 2420 to 5080 eV~cm$^{-3}$ in the CND.  

%
\begin{table*}
\begin{minipage}{\textwidth}
\centering
\scriptsize
\caption{Energy Density Distribution in Galaxies}
\begin{tabular}{lcccccc}
\hline
 & & Average Gas & CR & Radiation Field & Magnetic Field & Magnetic Field\\
 & Data & Density & Energy Density & Energy Density & Energy Density & Strength \\
 & & (cm$^{-3}$) & (eV~cm$^{-3}$) & (eV~cm$^{-3}$) & (eV~cm$^{-3}$) & ($\mu$G) \\
\hline
Milky Way & -- & 1 & 1.4 & 0.3 & 0.9 & 6\\
M31 & $\gamma$-rays & 1 & 1.52 & $\sim 0.5$ & 1.22 & 7\\
LMC & $\gamma$-rays & 2 & 0.58 & 0.3 & 0.4 & 4\\
\hline
\multirow{3}{*}{M82} & radio, $\gamma$-rays & \multirow{3}{*}{260} & 430 -- 620 & \multirow{3}{*}{480} & 1550 -- 3040 & 250 -- 350 \\
 & radio & & 280 -- 820 & & 990 -- 6210 & 200 -- 500 \\
 & $\gamma$-rays & & 550 & & -- & -- \\
\hline
\multirow{2}{*}{NGC 253} & radio, $\gamma$-rays & \multirow{2}{*}{680} & 260 -- 350 & \multirow{2}{*}{590} & 990 -- 1550 & 200 -- 250 \\
 & radio & & 130 -- 1290 & & 250 -- 6210 & 100 -- 500 \\
\hline
Arp 220 East & radio & 7700 & 1080 -- 4520 & 40 000 & $(0.4 - 1.4) \times 10^{6}$ & 4000 -- 7500 \\
Arp 220 ST & radio & 2810 & 1320 -- 4110 & 27 000 & $(0.25 - 1.6) \times 10^{5}$ & 1000 -- 2500 \\
Arp 220 CND & radio & 42 000 & 2420 -- 5080 & 440 000 & $(0.3 - 1.9) \times 10^{6}$ & 3500 -- 8750 \\
\hline
\multicolumn{7}{l}{Notes: Only the central starburst regions are considered for M82 and NGC 253. Uncertainities in the measurements of the}\\
\multicolumn{7}{l}{supernova rates are $\sim 50\%$, which is consistent with our results for both $U_{B}$ and $U_{\text{CR}}$ for M82 and NGC 253. As we have}\\
\multicolumn{7}{l}{no $\gamma$-ray data for Arp 220, our results are more unconstrained. Values for the Milky Way are taken from \citet{Ferriere01};}\\
\multicolumn{7}{l}{values for M31 \& the LMC are from \citet{Abdo10a}, \citet{Abdo10b}, \citet{Mao12} and references therein.}\\
\end{tabular}
\end{minipage}
\end{table*}
%

\subsubsection{Star-Forming Galaxies: Milky Way, M31, and LMC}

Aside from M82 and NGC~253, the best-studied star-forming galaxies detected in $\gamma$-rays include the Milky Way, M31, and the Large Magellanic Cloud.  We cannot apply the usual YEGZ models to constrain the energy densities, because diffusion is a critical process for CRs in these systems.  However, we can use the observed $\gamma$-ray flux to constrain the CR proton spectrum and to make a rough estimate of $U_{\text{CR}}$.

Assuming a power-law spectrum for the CR protons ($N_{p}(E) = A E^{-\Gamma}$), the observed $\gamma$-ray flux is given by
\begin{equation}
E_{\gamma}^{2} \frac{dN}{dE} = \frac{V}{4 \pi D^{2}} \frac{2 c n_{\text{ISM}}}{K_{\pi}} \int_{E_{\text{min}}}^{\infty} dE_{\pi} \frac{\sigma_{pp} N_{p}(m_{p}c^{2} - E_{\pi}/K_{\pi})}{\sqrt{E_{\pi}^{2} - m_{\pi}^{2}c^{4}}},
\end{equation}
where $E_{\text{min}} = E_{\gamma} + m_{\pi}^{2}c^{4} / 4E_{\gamma}$ and $K_{\pi} \approx 0.17$ is the fraction of kinetic energy transferred from the primary CR proton to the secondary pion.  For a given $\gamma$-ray flux and energy and an assumed gas density and volume (see Tables 1, 2), we can solve this equation for the normalization constant $A$ and the CR proton spectrum.  For M31 and the LMC, we find similar values for $U_{\text{CR}}$ of 1.52 and 0.58 eV~cm$^{-3}$, respectively.  We also estimated the CR energy density for M82, as a check of the method, and calculated an energy density of 550 eV~cm$^{-3}$, which is within the range of our results from the YEGZ models.

\subsection{Intense Star Formation and Equipartition?}

Taking the ratio of $U_{B}$ to $U_{\text{CR}}$ from our models, M82 diverges from equipartition by a factor of 2.5 to 7 and NGC 253 by a factor of 2.9 to 5.9 (see Figure 2, Table 2).  While these values are larger than originally estimated by \citet{Thompson06}, they are in agreement with calculations in \citet{Lacki14}.  Even more extreme are the nuclei of Arp 220.  The eastern nucleus of Arp 220 lies off of the equipartition relation by roughly one to three orders of magnitude, and the western nucleus lies off of the relation by factors of 60 to 785 in the CND and 6.0 to 52 in the surrounding torus.

The larger range in the possible values of acceleration efficiency and $U_{\text{CR}}$ in Arp 220 when compared to M82 and NGC~253 is due to the lack of $\gamma$-ray observations for the nuclei in Arp 220.  Because we are only comparing our models with the observed radio spectra of the nuclei, we find acceptable fits for every wind speed tested and for every combination of acceleration efficiency and spectral index.  If we disregard the $\gamma$-ray observations, then our simple models for M82 and NGC~253 also allow for a large range in the possible values for acceleration efficiency and $U_{\text{CR}}$ (see Figure 2, Table 2). 

We also find acceptable models where equipartition is upheld in fitting the radio spectra of M82 and NGC~253.  For those models, our best-fitting values for  magnetic field strengths and $U_{\text{CR}}$ are in rough agreement with those derived from the revised equipartition formula in \citet{Lacki13a}.  This further demonstrates the powerful impact of the $\gamma$-ray spectrum when constraining the total CR population and associated energy densities \citep[see also][]{Pohl93,Pohl94}.  Additionally, our models show a degeneracy for best fits to radio spectra such that as magnetic field strength increases, wind speed must also increase to achieve a similar goodness of fit.  As the wind speed is directly tied to the CR proton lifetime and thus the total CR proton spectrum, this means that as $U_{B}$ increases, $U_{\text{CR}}$ decreases, moving a given system further and further from equipartition.

Testing a physical range of magnetic field strengths, our best-fitting models lie off of equiparition.  An alternative explanation for the observed radio and $\gamma$-ray fluxes, which preserves equiparition, could be an increase in the number of primary electrons above the distribution observed in the Milky Way.  For M82, this requires a primary p/e ratio of 11 to 1.  While interesting, the astrophysical evidence for such a large change in this ratio is lacking \citep[e.g.,][]{Thompson09}.  However, this approach does not work in Arp 220, where we found that an equiparition magnetic field strength requires an even more extreme and unphysical p/e ratio of greater than 1 to 1, in favor of the electrons.

\section{Discussion and Conclusions}

In comparing our results for $U_{\text{CR}}$ and $U_{B}$, we naturally find that equipartition holds for the disks of the normal star-forming galaxies tested: the Milky Way, M31, and the LMC.  In contrast, while models derived only from radio continuum measurements agree with equipartition, the combined analysis of the radio and $\gamma$-ray data for M82 and NGC 253, with the normal electron to proton ratio, does not allow equipartition solutions.  As expected, due to inverse Compton losses \citep[e.g.,][]{Lisenfeld96, Thompson06} and particle advection in galactic winds \citep[e.g., YEGZ;][]{Domingo05}, in agreement with other studies we find that $U_{B}$ must be well above the minimum energy / equipartition value to produce the observed synchrotron luminosities in starbursts. The existence of the far-infrared radio correlation therefore indicates that intense magnetic fields must be routinely generated in zones of intense star formation \citep[see also][]{Schleicher13}. We expect this situation to prevail in any compact systems containing a dense ISM in combination with a high SFR, including many of the CMZs in nearby galaxies and high redshift galaxies where intense star formation and strong galactic winds are common. 

When considering the possible trends across all galaxies, we also see that both $U_{B}$ and $U_{\text{CR}}$ increase with increasing average gas density (see Figure 2).  Additionally, the ratio between $U_{B}$ and $U_{\text{CR}}$ increases with increasing average gas density.  Based on the Schmidt law, \citet{Lacki10} assumes $U_{CR} \propto \Sigma_{g}^{0.4}$ and $U_{B} \propto \Sigma_{g}^{1.4}$, which gives a ratio of $U_{B} / U_{CR} \propto n_{ISM}$.  Our results, based on densities derived from molecular line observations, show an upward trend in $U_{B} / U_{\text{CR}}$ versus $n_{\text{ISM}}$ (see Figure 2).  While the slope of this trend is in rough agreement with the relationship between energy densities and gas surface density assumed in \citet{Lacki10}, the error bars on our results for the nuclei of Arp 220 are extremely large.  So, depending on the constraints imposed by a future $\gamma$-ray detection, the correlation between $U_{B} / U_{CR}$ and the average gas density may have a shallower slope.

Similar to the assumptions of \citet{Lacki10}, we  find that $U_{\text{rad}}$ correlates with  $U_{B}$ but is systematically offset to lower values by a factor of $\sim 2$ to 5 for all galaxies tested \citep[see Equation 18 from][]{Murphy09}, with the possible exception of the eastern nucleus of Arp 220 (see Figure 2).  This close correlation between the magnetic and radiation fields is critical for the far-infrared-radio relationship to hold and indicates that the magnetic field strength is tied directly to SFRs, e.g., as discussed in \citet{Lacki13b}. This trend is consistent with models where $U_{B}$ is set by turbulence in the ISM which in turn depends on the intensity of star formation and structure of the system \citep[e.g.,][]{Balsara04,Thompson06,Lacki13a}. 

We have shown that equipartition is an unreliable assumption for CMZs.  However, we have only completed full analyses of three starburst galaxies.  Definitive establishment of a correlation between the magnetic field and CR energy densities and between the ratio $U_{B} / U_{CR}$ and the average gas density or the star-formation rate will benefit from further independent assessments of the energy densities in more galaxies.  Even simple models that jointly fit the radio and $\gamma$-ray spectra can offer powerful insights into feedback in star forming systems \citep[e.g.,][]{Ackermann12}.  Thus, more $\gamma$-ray detections of star-forming galaxies by \textit{Fermi} or other observatories and accompanying radio observations of the relevant regions will foster further progress in measuring $U_{B}$ and $U_{CR}$.  

Further, it is evident that zones with intense star formation, including the CMZs of nearby galaxies, may not be in equilibrium, as is often implicitly assumed.  CMZs therefore are potential laboratories for investigations of non-equilibrium astrophysical feedback on galactic scales. However, to fully utilize the potential of CMZs, we need to better understand the properties of CMZs, including multi-messenger studies \citep[see][]{Torres12, Gallagher14}.

Equipartition is sometimes also assumed to hold for galaxies at high redshift \citep[e.g.,][]{Schober15}.  This preliminary study  suggests that equipartition likely applies and can be a useful tool for determining magnetic field strengths for regions where star formation is spread out over a large area so that diffusive transport sets the dominant timescale for the CRs. Thus, equipartition may still be a useful tool at high redshift provided the star formation volume density is relatively low and there is careful consideration of the environment and associated timescales. However, for systems with compact volumes of star formation where advective and collisional energy losses set the CR lifetime, equipartition may not apply and caution is required in determining the magnetic field strength and other ISM energy densities.  This situation would apply, for example, in extreme starburst clumps and in compact galaxies with SFR intensities equal to or exceeding those in the M82 starburst, or $\sim 100 ~ M_{\odot}$~yr$^{-1}$~kpc$^{-2}$.

\section*{Acknowledgements}

This work was supported in part by NSF AST-0907837, NSF PHY-0821899 (to the Center for Magnetic Self-Organization in Laboratory and Astrophysical Plasmas), and NSF PHY-0969061 (to the IceCube Collaboration) and UW-Madison.  We thank the referee for their valuable comments and Francis Halzen for his help and support.


%
\end{document}